\documentclass[12pt]{article}
\def\half{\mbox{$\frac{1}{2}$}}

\title{Quantizing preserving Noether symmetries}
\author{M. C. Nucci}
\date{Dipartimento di Matematica
 e Informatica, Universit\`a di Perugia \& INFN Sezione Perugia, 06123 Perugia,
 Italy}
\begin{document}
\maketitle

\begin{abstract} A procedure which obviates the constraint imposed by the conflict between
consistent quantization and the invariance of the Hamiltonian description under
nonlinear canonical transformation is proposed. This new quantization scheme
preserves the Noether point symmetries of the underlying Lagrangian in order to
construct the Schr\"odinger's equation. As an example, the quantization of the
`goldfish' many-body problem  extensively studied by Calogero et al. is
presented.
\end{abstract}

\section{Introduction}

It has been known for over fifty years that quantization and nonlinear
canonical transformations have no guarantee of consistency \cite{van Hove 51
a}.  As recently stated by Brodlie in \cite{BrodlieJMP04} there is a
never-ending interest about ``the passage of canonical transformations from
classical mechanics to quantum mechanics". In \cite{Anderson94}, and reiterated
 in \cite{BrodlieJMP04}, it was said that {\em canonical transformations have three
important roles in both quantum and classical mechanics:
\begin{description}
\item{(i)} time evolution \item{(ii)} physical equivalence of two theories, and
\item{(iii)} solving a system. \end{description}}

  In this paper we propose a procedure which obviates the constraint
imposed by the conflict between consistent quantization and the invariance of
the Hamiltonian description under nonlinear canonical transformation. As far as
we know nobody has ever thought of a quantization scheme that preserves the
Noether point symmetries of the underlying Lagrangian in order to construct the
Schr\"odinger's equation.

In \cite {Goldstein 80 a} [ex. 18, p. 433] an alternative hamiltonian for the
simple harmonic oscillator was presented. It is obtained by applying a
nonlinear canonical transformation to the classical Hamiltonian of the harmonic
oscillator. In \cite{LGQM} that alternative Hamiltonian
 was used to demonstrate what nonsense the usual quantization
schemes produce. In \cite{gallipoli10} the quantization scheme that preserves
the Noether symmetries was applied  to the same example in order to derive the
correct Schr\"odinger's equation for the alternative Hamiltonian. We have
already inferred that Lie symmetries should be preserved if a consistent
quantization is desired \cite{c-iso}.

Our method  quantizes nonlinear Lagrangian equations -- i.e., any system of
equations that comes from a variational principle with a Lagrangian of first
order --
\begin{equation}
\ddot {\underline{x}}=\underline{f}(\underline{x},\dot{\underline{x}})
\end{equation}
that can be linearized through nonlinear canonical transformations.\\
 It yields the  Schr\"odinger's equation and can be summarized as follows
\begin{enumerate}
\item Find the Lie symmetries of the Lagrange equations
$$\Upsilon=W(t,\underline{x})\partial_t+\sum_{k=1}^{N}W_k(t,\underline{x})\partial_{x_k}$$
\item  Among them find the Noether's symmetries
$$\Gamma=V(t,\underline{x})\partial_t+\sum_{k=1}^{N}V_k(t,\underline{x})\partial_{x_k}$$
This may require to look for the Lagrangian yielding the maximum possible
number of Noether's symmetries \cite{laggal}, \cite{CP07Rao1JMP},
\cite{nuctam_1lag}, \cite{nuctam_3lag}
\item  Construct the Schr\"odinger's equation admitting these Noether's
symmetries as Lie symmetries
$$2iu_t+\sum_{k,j=1}^{N} f_{kj}(\underline{x})u_{x_jx_k}+
\sum_{k=1}^{N}h_k(\underline{x})u_{x_k}+f_3(\underline{x})u=0$$
$$\Omega=V(t,\underline{x})\partial_t+\sum_{k=1}^{N}V_k(t,\underline{x})\partial_{x_k}
+G(t,\underline{x},u)\partial_u$$ without adding any other symmetries apart the
two symmetries that are present in any linear partial differential equation,
namely
$$u\partial_u, \quad \quad \alpha(t,\underline{x})\partial_u,$$
where $\alpha=\alpha(t,\underline{x})$ is any solution of the Schr\"odinger's
equation.
\end{enumerate}

As an example, the quantization of the `goldfish' many-body problem extensively
studied by Calogero et al is presented: in Section 2 we find the Lie and
Noether's symmetries of the two-body problem; in Section 3 we derive the
 Schr\"odinger's equation of the two-body problem and the
general formula yielding the Schr\"odinger's equation of the
`goldfish' many-body problem.\\

 In \cite{calo78} Calogero  derived  a solvable many-body problem,
i.e.
\begin{equation}
\ddot x_n=2\sum_{m=1,\;m\neq n}^{N}{\dot x_n \dot x_m \over x_n-x_m}
,\;\;\;\;(n=1,\ldots, N) \label{gf}
\end{equation}
by considering  the following solvable nonlinear partial differential equation:
$$ \varphi_t+\varphi_x+\varphi^2=0,
\;\;\;\;\quad\varphi\equiv\varphi(x,t)$$ and looking at the behavior of the
poles of its solution. In \cite{calo01} the same system (\ref{gf})  was
presented, its properties were further studied and its solution was given in
terms of the roots of the following algebraic equation in $x$:
\begin{equation}
\sum_{m=1}^{N}{\dot x_m(0)\over[x-x_m(0)]}={1\over t} \label{calform}
\end{equation}
In that paper, Calogero called system (\ref{gf}) ``a goldfish" following a
statement by Zakharov \cite{zakh} [p. 622], namely {\em A mathematician, using
the dressing method to find a new integrable system, could be compared with a
fisherman, plunging his net into the sea. He does not know what a fish he will
pull out. He hopes to catch a goldfish, of course. But too often his catch is
something that could not be used for any known purpose to him. He invents more
and more sophisticated nets and equipments, and plunges all that deeper and
deeper. As a result, he pulls on the shore after a hard work more and more
strange creatures. He should not despair, nevertheless. The strange creatures
may be interesting enough if you are not too pragmatic, and who knows
how deep in the sea do goldfishes live?}\\
Calogero and others have extensively studied system (\ref{gf}), e.g.
\cite{calobook}, \cite{gf}, \cite{GomSom05}, \cite{Guillot05}.

\section{Lie and Noether's symmetries of the ``goldfish" two-body problem}
In the case $N=2$ system (\ref{gf}) reduces to
\begin{eqnarray}
\ddot x_1&=&2{\dot x_1\dot x_2\over x_1-x_2}\nonumber\\
\ddot x_2&=&-2{\dot x_1\dot x_2\over x_1-x_2}. \label{N2gf}
\end{eqnarray}
Using the interactive REDUCE programs \cite{man}, we obtain a
fifteen-dimensional Lie point symmetry algebra -- that is isomorphic to
$sl(4,I\!\!R)$ \cite{gasgon}, \cite{gonzalez88} -- generated by the following
fifteen operators:
\begin{eqnarray}
\Gamma_1&=&\frac{x_1x_2}{x_1-x_2}\left(t(x_1-x_2)\partial_t+x_1^2\partial_{x_1}-
x_2^2\partial_{x_2}\right)\nonumber \\
\Gamma_2&=&x_1x_2\partial_t\nonumber \\
\Gamma_3&=&t(x_1+x_2)\partial_t+x_1^2\partial_{x_1}+
x_2^2\partial_{x_2}\nonumber \\
\Gamma_4&=&(x_1+x_2)\partial_t\nonumber \\
\Gamma_5&=& -\frac{x_1x_2}{x_1-x_2}\left(x_1\partial_{x_1}-
x_2\partial_{x_2}\right)\nonumber \\
\Gamma_6&=&
\frac{1}{2(x_1-x_2)}\left(2t(x_1-x_2)\partial_t+x_1^2\partial_{x_1}-
x_2^2\partial_{x_2}\right)\nonumber \\
\Gamma_7&=& \partial_t\nonumber \\
\Gamma_8&=& -\frac{t}{x_1-x_2}\left(x_1\partial_{x_1}-
x_2\partial_{x_2}\right)\nonumber \\
\Gamma_9&=& -\frac{1}{x_1-x_2}\left(x_1\partial_{x_1}-
x_2\partial_{x_2}\right)\nonumber \\
\Gamma_{10}&=& -\frac{t}{x_1-x_2}\left(\partial_{x_1}-
\partial_{x_2}\right)\nonumber \\
\Gamma_{11}&=& -\frac{1}{x_1-x_2}\left(\partial_{x_1}-
\partial_{x_2}\right)\nonumber \\
\Gamma_{12}&=& \frac{t}{x_1-x_2}\left(t(x_1-x_2)\partial_t+x_1^2\partial_{x_1}-
x_2^2\partial_{x_2}\right)\nonumber \\
\Gamma_{13}&=& -\frac{1}{3}\left(x_1\partial_{x_1}+
x_2\partial_{x_2}\right)\nonumber \\
\Gamma_{14}&=& -\frac{1}{3(x_1-x_2)}\Big(\left(2x_1+x_2\right)\partial_{x_1}-
(x_1+2x_2)\partial_{x_2}\Big)\nonumber \\
\Gamma_{15}&=&-\frac{1}{3(x_1-x_2)}\left((x_1^2+2x_1x_2)\partial_{x_1}-
(x_2^2+2x_1x_2)\partial_{x_2}\right) \label{gammas}
\end{eqnarray}
which means that system (\ref{N2gf})
 is linearizable \cite{fels}, \cite{sohmaho01}.
 In order to find the linearising transformation we look for a four-dimensional
  abelian subalgebra $L_{4,2}$ of rank 1
 and  have to transform it into the canonical form \cite{sohmaho01}
$$\partial_{y},\;\;\; y_1 \partial_{ y},\;\;\;  y_2\partial_{
y},\;\;\; y\partial_{ y}, $$ with $ y$, $ y_1$ and $y_2$ the new independent
and dependent
 variables, respectively.
 We find that one such
subalgebra
  is that generated by
  \begin{equation}
\Gamma_7=\partial_t,\quad\Gamma_4=(x_1+x_2)\partial_t,\quad\Gamma_2=x_1x_2\partial_t,
\quad\Gamma_6-\Gamma_3-\half\Gamma_{15}=t\partial_t\,.
\end{equation} Then, it is
easy to derive that the linearizing transformation is
 \begin{equation} y=t, \quad\quad y_1=x_1+x_2, \quad\quad
y_2=x_1x_2 \label{cantr} \end{equation}
  and system (\ref{N2gf})
becomes
\begin{equation}
\ddot y_1= 0, \quad\quad \ddot y_2= 0 \label{free2eq}
\end{equation}
which may be interpreted as the equations of motion of a free particle on a
plane.

The hidden linearity of system (\ref{N2gf}) is already known \cite{calo01}.\\\\
Since the kinetic energy of a free particle on a plane is
\begin{equation}T=\half(\dot y_1^2+\dot y_2^2),\end{equation}
then transformation (\ref{cantr}) yields the following Lagrangian for system
(\ref{N2gf}):
\begin{equation}
L=\half \left ((\dot x_1+\dot x_2)^2+(x_2\dot x_1 +x_1\dot
x_2)^2\right)+\frac{{\rm d}g}{{\rm d}t}\,,\label{N2lag}
\end{equation}
where $g=g(t,x_1,x_2)$ is the gauge function, a fundamental element if one
wants to apply Noether's theorem correctly.

This Lagrangian admits eight Noether point symmetries \cite{gonzalez88} out of
the fifteen Lie point symmetries in (\ref{gammas}), i.e.:
\begin{eqnarray}
\Gamma_5+3\Gamma_{14}&=&
-\frac{1}{x_1-x_2}\left((x_1^2x_2+2x_1+x_2)\partial_{x_1}-
(x_1x_2^2+x_1+2x_2)\partial_{x_2}\right)\nonumber \\
\Gamma_6&=&
\frac{1}{2(x_1-x_2)}\left(2t(x_1-x_2)\partial_t+x_1^2\partial_{x_1}-
x_2^2\partial_{x_2}\right)\nonumber \\
\Gamma_7&=& \partial_t\nonumber \\
\Gamma_8&=& -\frac{t}{x_1-x_2}\left(x_1\partial_{x_1}-
x_2\partial_{x_2}\right)\nonumber \\
\Gamma_9&=& -\frac{1}{x_1-x_2}\left(x_1\partial_{x_1}-
x_2\partial_{x_2}\right)\nonumber \\
\Gamma_{10}&=& -\frac{t}{x_1-x_2}\left(\partial_{x_1}-
\partial_{x_2}\right)\nonumber \\
\Gamma_{11}&=& -\frac{1}{x_1-x_2}\left(\partial_{x_1}-
\partial_{x_2}\right)\nonumber \\
\Gamma_{12}&=& \frac{t}{x_1-x_2}\left(t(x_1-x_2)\partial_t+x_1^2\partial_{x_1}-
x_2^2\partial_{x_2}\right)
\end{eqnarray}
To each Noether's symmetry corresponds a first integral of system (\ref{N2gf}).
For example $\Gamma_7$ yields the Lagrangian (\ref{N2lag}) itself as a
conserved quantity.\\

It was proven in \cite{gonzalez88} that the  $n^2+4n+3$-dimensional (i.e., of
maximal dimension) Lie symmetry algebra of a system of $n$ equations of second
order is isomorphic to $sl(n+2,I\!\!R)$, and the corresponding Noether
symmetries generate a $(n^2+3n+6)/2$-dimensional Lie algebra $g^V$ whose
structure (Levi-Mal\'cev decomposition and realization by means of a matrix
algebra) was determined. Recently  the Lie and Noether symmetries of a non
autonomous linear Lagrangian system of two second-order equations, i.e.
\begin{equation}
\ddot q_1=-\frac{k}{m}q_1+\frac{t}{m}, \quad\quad \ddot
q_2=-\frac{k}{m}q_2.\label{eqFu}
\end{equation}
 were
determined \cite{oldconsqPLA}.

\section{Quantization of the ``goldfish"}
The  Hamiltonian corresponding to the Lagrangian (\ref{N2lag}) is:
\begin{equation}
H=\frac{1}{2(x_1-x_2)^2}\left((p_1x_1-p_2x_2)^2+(p_1-p_2)^2\right).
\end{equation}

One may try to quantize this Hamiltonian by using the various classical
methods. Neither the normal ordering method nor the Weyl quantisation procedure
lead to a result which is physical \cite{LGQM}. This is due to the nonlinearity
of the canonical transformation (\ref{cantr}) between system (\ref{N2gf}) and
system (\ref{free2eq}).

Instead we assume that the Schr\"odinger's equation corresponding to system
(\ref{N2gf}) be of the following type:
\begin{equation}
2iu_t+\sum_{k,j=1}^{2} f_{kj}(x_1,x_2)u_{x_jx_k}+
\sum_{k=1}^{2}h_k(x_1,x_2)u_{x_k}+h_0(x_1,x_2)u=0 \label{N2gfschp}
\end{equation}
with $f_{kj},h_k,h_0$ functions of $x_1,x_2$ to be determined in such a way
that equation (\ref{N2gfschp}) admits the following eight Lie symmetries:
\begin{eqnarray}
\Gamma_5+3\Gamma_{14}&\Rightarrow&\Omega_1 =
-\frac{1}{x_1-x_2}\left((x_1^2x_2+2x_1+x_2)\partial_{x_1}-
(x_1x_2^2+x_1+2x_2)\partial_{x_2}\right)+ \omega_1\partial_u\nonumber \\
\Gamma_6&\Rightarrow&\Omega_2=
\frac{1}{2(x_1-x_2)}\left(2t(x_1-x_2)\partial_t+x_1^2\partial_{x_1}-
x_2^2\partial_{x_2}\right)+ \omega_2\partial_u\nonumber \\
\Gamma_7&\Rightarrow& \Omega_3= \partial_t+ \omega_3\partial_u\nonumber \\
\Gamma_8&\Rightarrow&\Omega_4= -\frac{t}{x_1-x_2}\left(x_1\partial_{x_1}-
x_2\partial_{x_2}\right)+ \omega_4\partial_u\nonumber \\
\Gamma_9&\Rightarrow& \Omega_5= -\frac{1}{x_1-x_2}\left(x_1\partial_{x_1}-
x_2\partial_{x_2}\right)+ \omega_5\partial_u\nonumber \\
\Gamma_{10}&\Rightarrow& \Omega_6= -\frac{t}{x_1-x_2}\left(\partial_{x_1}-
\partial_{x_2}\right)+ \omega_6\partial_u\nonumber \\
\Gamma_{11}&\Rightarrow& \Omega_7= -\frac{1}{x_1-x_2}\left(\partial_{x_1}-
\partial_{x_2}\right)+ \omega_7\partial_u\nonumber \\
\Gamma_{12}&\Rightarrow& \Omega_8=
\frac{t}{x_1-x_2}\left(t(x_1-x_2)\partial_t+x_1^2\partial_{x_1}-
x_2^2\partial_{x_2}\right)+ \omega_8\partial_u
\end{eqnarray}
where $\omega_i=\omega_i(t,x_1,x_2,u), (i=1,8)$ are functions of $t,x_1,x_2,u$
that have to be determined. Equation (\ref{N2gfschp}) also admits the following
two symmetries
\begin{equation}
\Omega_9=u\partial_u, \quad\quad\quad
\Omega_{\alpha}=\alpha(t,x_1,x_2)\partial_u \label{twosym}
\end{equation}
with $\alpha$ any solution of equation (\ref{N2gfschp}) itself, since any
linear partial differential equation possesses these  two symmetries.

Using the interactive REDUCE programs \cite{man}, we obtain that
\begin{eqnarray}
f_{11}=\frac{x_1^2 + 1}{(x_1 - x_2)^2},\quad f_{12}=f_{21}=-\frac{x_1x_2 +
1}{(x_1 - x_2)^2},\quad f_{22}=\frac{x_2^2 + 1}{(x_1 - x_2)^2},\nonumber\\
h_1=\frac{\partial f_{11}}{\partial x_1},\quad h_2=\frac{\partial
f_{22}}{\partial x_2},\quad h_0=-E_0^2
\end{eqnarray}
and
\begin{eqnarray}
\omega_1=0,\quad  \omega_2=-\half iE_0^2 t u, \quad \omega_3=0, \quad
\omega_4=-i(x_1+x_2)u, \quad \omega_5=0, \nonumber \\ \omega_6=ix_1x_2 u, \quad
\omega_7=0,\quad \omega_8=(i
 x_1x_2-t)u+\frac{i u}{2}\left(x_1^2x_2^2+x_1^2+x_2^2
 -t^2E_0^2\right),
\end{eqnarray}
with $E_0$ an arbitrary constant. Therefore the Schr\"odinger's equation of
system (\ref{N2gf}) is
\begin{eqnarray}
2iu_t+\frac{x_1^2 + 1}{(x_1 - x_2)^2}\,u_{x_1x_1}- 2\frac{x_1x_2 + 1}{(x_1 -
x_2)^2}\, u_{x_1x_2}+\frac{x_2^2 + 1}{(x_1 - x_2)^2}\, u_{x_2x_2}\nonumber
\\+\frac{\partial}{\partial x_1}\left(\frac{x_1^2 + 1}{(x_1 -
x_2)^2}\right)u_{x_1}+\frac{\partial}{\partial x_2}\left(\frac{x_2^2 + 1}{(x_1
- x_2)^2}\right)u_{x_2}-E_0^2u=0. \label{sch2Ngf}
\end{eqnarray}
In fact if we assume $u=\psi(t,y_1,y_2)$ with $y_1=x_1+x_2, y_2=x_1 x_2$ as
given in (\ref{cantr}) then equation (\ref{sch2Ngf}) becomes the well-known
Schr\"odinger's equation for the two-dimensional free particle, i.e.:
\begin{equation}
2i\psi_t+\psi_{y_1y_1}+\psi_{y_2y_2}-E_0^2\psi=0.
\end{equation}
It is now obvious that if the Schr\"odinger's equation for the N-dimensional
free particle is considered, i.e.
\begin{equation}
2i\psi_t(t,\mathbf{y})+\triangle
\psi(t,\mathbf{y})-E_0^2\psi(t,\mathbf{y})=0,\quad \mathbf{y}=(y_1,\ldots,y_N),
\end{equation}
then the transformation
\begin{equation} u=\psi(t,\mathbf{y}), \quad \mathbf{y}=\left(\sum_{i=1}^{N}x_i,
 \sum_{{i,j=1,\;i<j}}^N x_ix_j,\sum_{{i,j,k=1,\;i<j<k}}^N x_ix_jx_k,\,
 \ldots\,,\prod_{i=1}^{N}x_i\right) \end{equation}
 yields the Schr\"odinger's equation of system (\ref{gf}).

\end{document}